\documentclass[nofootinbib,aps,prd,preprint,groupedaddress]{revtex4-1}

\usepackage{graphicx}% Include figure files
\usepackage{dcolumn}% Align table columns on decimal point
\usepackage{bm}% bold math
\usepackage{axodraw}
\usepackage{amssymb}
\usepackage{amsmath,amsthm,amsfonts} 
\usepackage{bigints}
\usepackage{color}

\def\R{\mathbb{R}}

\DeclareMathOperator*{\under}{\sim} 

\begin{document}

\preprint{MaPhy-AvH/2013-01}

\title{Resummation of infrared logarithms in de Sitter space via Dyson--Schwinger equations: the ladder--rainbow approximation}

\author{A. Youssef}
\email{youssef@mathematik.hu-berlin.de}
\author{D. Kreimer}
\email{kreimer@mathematik.hu-berlin.de}
\affiliation{Institut f\"ur Mathematik und Institut f\"ur Physik, Humboldt-Universit\"at zu Berlin\\ Johann von Neumann-Haus, Rudower Chaussee 25, 12489 Berlin, Germany}
\date{\today}

\begin{abstract}
We study the infrared (large separation) behaviour of a massless minimally coupled scalar QFT with a quartic self--interaction in de Sitter. We show that the perturbation series in the interaction strength is singular and secular, i.e it does not lead to a uniform approximation of the solution in the infrared region. Only a non--perturbative resummation can capture the correct infrared behaviour.  We seek to justify this picture using the Dyson--Schwinger equations in the ladder--rainbow approximation. We are able to write down an ordinary differential equation obeyed by the two--point function and perform its asymptotic analysis.  Indeed, while the perturbative series -- truncated at any finite order -- is growing in the infrared, the full non-perturbative sum can be decaying.\end{abstract}


\maketitle

\section{Introduction}
 
Questions on the actual size of quantum loops corrections during inflation \cite{Senatore:2010ec,Senatore:2013ux,Gerstenlauer:2011vp}, the eventual dynamical screening of the cosmological constant and the related instability of de Sitter spacetime \cite{Polyakov:1982dz,Polyakov:2007mm,Jackiw:2005wk} are of great interest to cosmology. As part of an ongoing effort to illuminate these difficult questions, interacting QFT in de Sitter spacetime has been attracting a lot of attention recently. Since the literature on the subject can easily become confusing, we refer the interested reader to the recent and much needed review \cite{Akhmedov:2014ws}.

\medskip

In this paper we will concentrate on a  specific and important aspect of interacting QFT in de Sitter spacetime, namely \textit{the infrared behaviour of the massless self--interacting scalar field}. We will not discuss the back--reaction induced by the stress--energy tensor of the latter on the de Sitter background, although our results are of direct relevance to this matter. We also note that, while the major conclusions of our work concerning the IR behaviour of interacting fields in de Sitter are likely to be relevant to more general cosmological models, this relation is certainly not immediate, due to the special form of the $\lambda \phi^4$ interaction we consider.

\medskip
Beside of the usual vacuum ambiguities specific to curved spacetime field theories, the quantization of the free massive scalar field in de Sitter poses no particular challenges\footnote{See however the papers \cite{Polyakov:2007mm,Anderson:2013ila,Anderson:2013vy} for a very interesting line of research claiming that certain vacuum states, even for free/massive fields, might be themselves unstable.}. In the so--called Bunch--Davies vacuum state, the two--point function $\Delta_{m^2}$ is a perfectly well defined hypergeometric function of the invariant distance. For small masses however, $\Delta_{m^2}$ develops a  $1/m^2$ pole, rendering the massless limit ill-defined and the quantization of the  theory a non-trivial task. This situation must be contrasted with the flat space case, where the massless limit is smooth. We note that the divergence appearing in the limit $m\to0$ is often referred to as an --important-- IR divergency in de Sitter. As will become clear in what follows, we believe that this is neither the relevant notion of IR divergency, nor is it an important issue in the first place.

\medskip
Several propositions to quantize the massless field co-exist in the literature. Depending on how one interprets and cures the ill-defined massless limit, one might end up with a de Sitter invariant theory or not. 

The first and most popular option is to understand the problematic massless limit as an indication of the impossibility of maintaining the full de Sitter invariance of the quantum theory. Indeed, breaking de Sitter invariance leads to a well-defined quantization and to  interesting cosmological implications \cite{Allen:1987tz,Woodard:2004te}.

Another approach can be found in \cite{Folacci:1992cl,Bros:2010fu}, where a ``renormalized" two--point function $\Delta_0$ is defined through the subtraction of the divergent $1/m^2$ term. The resulting quantum theory preserves de Sitter invariance, as well as causality and positivity. We have also shown in \cite{Youssef:2012cx} that it leads to the observed scale-invariant CMB power spectrum. This ``renormalized" free propagator will be used throughout the present article. 

\medskip
A most important fact is that while the massive two--point function $\Delta_{m^2}$ decays in the IR region, $\Delta_0$ is growing in this same limit. This is the interesting IR notion we believe one should study. The rest of the article is devoted to the study of what becomes of this growing behaviour when a quartic self--interaction is added to the Lagrangian.   

\medskip
Most of the preceding works dealing with this problem focus on the simplest approximation possible, namely local contributions to the self--energy, either at a finite order in perturbation theory, or non--perturbatively by summing the infinite set of the so--called cactus diagrams. The outcome of such an approximation is at most a dynamically generated mass $m_\textrm{dyn}^2$. Indeed most of the authors considering this approximation find $m_\textrm{dyn}^2 \propto \sqrt{\lambda} H^2$-where $\lambda$ is the interaction strength and $H$ is the Hubble constant (see \cite{Kahya:2007dk,Burgess:2010ji,Serreau:2011gt,Garbrecht:2011gu,Boyanovsky:2012ix,Beneke:2013eh}
 and references therein). A non-vanishing $m_\textrm{dyn}^2$ is also what the earlier and more sophisticated Starobinsky's stochastic approach \cite{Starobinsky:1994is,Woodard:2007ft} leads to.

Unfortunately these results strongly depend on how one defines the coincidence limit of the free massless two-point function. Indeed, even at the much simpler flat space level, where no dynamical mass generation is believed to occur, the implications of the renormalization schemes have not been fully clarified, especially because renormalization schemes that are mathematically acceptable might not always be physically meaningful. Anyhow, an eventual dynamically generated mass in de Sitter would imply that the growing IR behaviour in de Sitter has been cured by the inclusion on an interaction, however mild. In other terms, the presence of a small but non-vanishing $\lambda$ will make the massless limit smooth and the limits $m\to0$ and $\lambda \to0$ do not commute. 

\medskip
In this paper go beyond this local approximation by considering non--local contributions to the self-energy. An effort to go beyond this approximation can also be found in \cite{Gautier:2013ws}, where the Dyson series of the two-loops self--energies are summed. We note that, contrary to the flat space case, the summation of the Dyson series of repeated self--energies chains is a non--trivial task in de Sitter space. Another attempt to consider general massless non--local diagrams has been made in \cite{Hollands:2011iy}, where interesting parametric representations of the interacting propagator have been developed.

A first reason to consider non--local contributions is the ambiguity of the local approximations discussed above. Perhaps the most important motivation for us is to prove in a well defined and specific context, that \textit{perturbation theory is singular in de Sitter spacetime}, and to insist that this is probably the single most important technical issue underlying all of de Sitter interacting QFT.

\medskip
Considering non-local diagrams is way more challenging technically speaking in de Sitter spacetime, and specially in the massless limit. We are essentially faced with two major difficulties. First of all the absence of translational symmetry makes the harmonic analysis much less useful in de Sitter calculations. This has to be contrasted with the extremely powerful  tool that is the flat space Fourier transform. We invite the readers to compare the relative difficulties of computing say a simple one--loop two--point function for a scalar massive field theory in de Minkowski (see for instance chapter 3 of \cite{Collins:1984vl}) and in de Sitter \cite{Marolf:2010cd}.

Second, one must add up the difficulties proper to the massless case. While in flat space this is a simplifying limit, in de Sitter, because the free massless propagator is growing in the IR, any Feynman integral using it as a building block is plagued with many IR divergencies.  As will become shortly clear, our setup solves very efficiently the two preceding difficulties by intensively making use of de the Sitter symmetry and by transforming  integral equations into differential ones.

\medskip
While we indeed believe that the inclusion of an interaction, however small,  has drastic consequences on the quantization of the massless field and its IR behaviour, we argue here in  favour of a more precise overall picture. The free propagator $\Delta_0$, as well as its perturbative (non--local) corrections are growing. However, when the leading IR contributions are summed up, the full non-perturbative propagator $G$ is decaying. In other words, the perturbative expansion develops secular terms, making the perturbative series non--uniform in the IR. We take a first step to justify this picture by summing, via a linear Dyson--Schwinger equation, the so called ladder--rainbow diagrams.

\medskip

The organization of the paper is as follows: After a rapid introduction to the basics of de Sitter geometry and QFT, we illustrate the methods we use on the simpler cases of massless flat space and massive de Sitter space fields. We then write down in section \ref{DSequations} the Dyson--Schwinger equations in the local and in the ladder--rainbow approximation. We give a rapid discussion on the local approximation and consequent dynamical mass generation. Taking full-advantage of de Sitter symmetry, we are able to reduce the  Dyson--Schwinger integral equation into an ordinary differential one, the asymptotics of which is studied in section \ref{asymptoticofmaster}. In section \ref{nonlinearladderrainbow} we merely write down the non-linear DS equations and suggest a strategy to tackle them in the future. Finally, section \ref{discussion} is a concluding discussion. 

\section{De Sitter QFT}

\subsection{De Sitter geometry}
The $D$-dimensional de Sitter spacetime can be identified with the real one-sheeted hyperboloid in the $D+1$ Minkowski spacetime $M_{D+1}$:
$$
X_D=\left\lbrace x \in \R^{D+1}, x^2=-H^{-2}\right\rbrace
$$
where $H>0$ is the Hubble constant. Flat space is obtained as the $H\to0$ limit.
This definition of the de Sitter manifold reveals the maximal symmetry of $X_D$ under the action of the de Sitter group $SO_0(1,D)$, the latter making the calculations analytically tractable. We take full advantage of this fact in this paper.  Let $\mu(x,x')$ denote the geodesic distance between the points $x$ and $x'$. It is useful to introduce the quantity $z$ given for spacelike separations ($\mu^2>0$ or $0<z<1$) by $z=\cos^2 \left( \frac{H \mu}{2}\right)$. As a general rule, we will work in the spacelike region and then analytically continue the propagators.
As shown in \cite{Allen:1986wq}, when acting on functions of the invariant $\mu(x,x')$, the Laplace-Beltrami operator reduces to the ordinary differential operator:
\begin{equation}
\label{box}
(-\Box+m^2)=-\frac{\mathrm{d}^2}{\mathrm{d}\mu^2}-(D-1)A(\mu) \frac{\mathrm{d}}{\mathrm{d}\mu}+m^2
\end{equation}
where 
\begin{equation*}
\label{functionA}
\left\{
\begin{array}{rll}
A(\mu)&=\mu^{-1},\quad &\textrm{on Minkowski space}\\
A(\mu)&=H \cot(H \mu),\quad &\textrm{on de Sitter space.}
\end{array} \right.
\end{equation*}  
In terms of the variable $z$, this becomes the hypergeometric operator
$$
(-\Box+m^2)=\mathbf{H}=-z(1-z)\frac{\mathrm{d}^2}{\mathrm{d}z^2}-\left[c-(a_{+}+a_{-}+1)z\right]\frac{\mathrm{d}}{\mathrm{d}z}+a_{+}a_{-}
$$
where
$$
a_{\pm}=\dfrac{D-1\pm\sqrt{(D-1)^2-\frac{4m^2}{H^2}}}{2}, \quad c=D/2.
$$
In particular, in the massless case we have
\begin{equation}
\label{operatorH}\mathbf{H}=-z(1-z)\frac{\mathrm{d}^2}{\mathrm{d}z^2}-\frac{D}{2}(1-2z)\frac{\mathrm{d}}{\mathrm{d}z}.
\end{equation}

In order to make the connection to cosmology more direct, we also introduce a flat coordinate system. The spatial sections in this coordinate system are $D-1$ planes. These coordinates cover only half of the de Sitter spacetime and are given by:
\begin{equation*}
x(t,\boldsymbol{x})= 
\left\{
 \begin{array}{rl}
  x^0 &= H^{-1}\sinh(H t)+\frac{H}{2} \boldsymbol{x}^2 e^{H t} \\
  x^j &= \boldsymbol{x}^j e^{H t} \\
  x^D &= H^{-1}\cosh(H t)-\frac{H}{2} \boldsymbol{x}^2 e^{H t} 
 \end{array} \right.
\end{equation*}

 The de Sitter metric and the invariant quantity $z$ in this coordinate system read
\begin{align*}
ds^2&=dt^2-a^2(t) d\boldsymbol{x}^2, \quad a(t)=e^{H t}\\
z(x,x')&=\frac{1}{2} \left[1-\frac{H^2}{2}e^{H(t+t')} (\boldsymbol{x}-\boldsymbol{x}')^2 +\cosh(H(t-t'))\right]
\end{align*}

\subsection{The massless field}
\label{}

The physical reason behind the appearance of strong IR effects in de Sitter can be simply understood: the rapid expansion of the spacetime dilate correlation patterns. After all this is the exact reason why a de Sitter inflationary phase in the early  universe solves many problems of the hot big-bang model. These IR effects are  naturally stronger for massless (and non-conformally invariant) fields  such as the massless minimally coupled (mmc) \footnote{Minimally coupled means that there is no term in the action proportional to $\mathcal{R} \phi^2$, $\mathcal{R}$ being the Ricci scalar. As the latter is constant in de Sitter, it is a correction to the mass term.} scalar and the graviton. We review here these IR divergences in the mmc case. 

Recall that in the Bunch--Davies vacuum state the massive propagator for a scalar field reads:

\begin{equation}
\label{massivefreesolution}
\Delta(x,x')=\frac{\Gamma(a_{-})\Gamma(a_{+}) H^{D-2}}{\Gamma(D/2) 2^D \pi^{D/2}}\;  {_{2}}F_{1}\left[a_{-},a_{+},D/2,z\right].
\end{equation}
In the massless limit, this expression diverges because of the pole of the Gamma function. More precisely we have the small mass expansion:
$$
\Delta(x,x') \sim \frac{D}{4 \pi^{\frac{D+1}{2}}} \Gamma\left(\frac{D-1}{2}\right)\frac{H^D}{m^2}+ \textrm{regular terms in $m$}.
$$ 
Note that in the flat space limit ($H\to 0$), this singular term is absent and the massless limit is smooth \footnote{\label{noncomm}This means that the flat space limit ($H\to 0$) and the massless limit  ($m\to 0$) do not commute. This is a physically important fact and might mean that even a small amount of curvature - like in today's universe - could have important consequences on massless fields.}. 

One of the first papers studying the mmc scalar field  in de Sitter is \cite{Allen:1987tz}, where the authors prove that a usual de Sitter-invariant Fock space quantization is impossible in this case. They then propose to trade the de Sitter SO$(1,D)$ invariance for a smaller one, say a SO$(D)$ invariance. Equivalently, it is a common belief among workers in the field that the  scale-invariant power-spectrum leads necessarily to a breakdown of de Sitter invariance and that some physical quantities might thus become time-dependent. Several authors later proposed different treatments of the mmc field, among which \cite{Bros:2010fu} is one of the most exhaustive. Here the divergent term is  subtracted, a ``renormalized'' de Sitter-invariant two--point function is computed \footnote{The draw-back  is that the two--point function no longer verifies the equation of motion $\Box \phi=0$, instead it verifies the anomalous equation:
$$
\Box \phi=-\frac{3}{8 \pi^{2}}.
$$ 
This simple renormalization procedure has been used implicitly in several earlier works. However, the major contribution of \cite{Bros:2010fu} is proving that on a suitably chosen subspace of states $\mathcal{E}$, the equation of motion is effectively restored. This ``Physical'' space of states should be regarded the same way as we regard the one that appears in the quantization of gauge theory (for instance the space of transverse photons in QED). Moreover, the authors were able to show that the renormalized two--point function defines a positive kernel when restricted to $\mathcal{E}$, thus enabling a probabilistic interpretation of the theory.} and it reads (we work in $D=4$ for simplicity):
\begin{equation}
\label{rw}
\Delta_0(z)=\frac{H^2}{(4\pi)^2} \left[ \frac{1}{1-z}-2 \ln (1-z)\right].
\end{equation}
In \cite{Youssef:2012cx} we also  proved that $\Delta_0$ gives, as it should, the observed scale-invariant power spectrum. We will use  this propagator for the mmc hereafter (and we will omit the subscript $0$ when no confusion can arise).

\section{Illustration of our approach}

In this article we use a calculation method that is not habitually used in de Sitter spacetime, moreover we use it to study the complicated massless field.  For the purpose of clarity, we try to  disentangle theses difficulties by illustrating  first our techniques on the flat space case. We then consider de Sitter spacetime but for the simpler massive field.

For simplicity, we will only consider in this section the cubic self-interacting scalar field in $6$ dimensional spacetime whose action is:
$$
S=\int dV_x \left[\frac{1}{2} g_{\mu\nu}\partial^\mu \phi \partial^\nu \phi-\frac{1}{2} m^2 \phi^2-\frac{\lambda}{3!}\phi^3 \right],
$$
where $dV_x=\sqrt{-g} d^6x$ is the invariant volume element.

\subsection{Flat spacetime} 
The position-space $D$-dimensional free two-point function is usually computed from the Fourier-space function and is known to be equal to (for space-like separations):
$$
\Delta=c_D \; \left(\frac{m^2}{\mu^2}\right)^{(D-1)/4} K_{\frac{D}{2}-1}\left(\sqrt{m^2\mu^2}\right)
$$
where $c_D$ is a numeric constant independent from $m$ and $K$ is the modified Bessel function of the second kind.  In the previous expression $\mu(x,x')=\sqrt{(x-x')^2}$ is the invariant distance. This result can be retrieved immediately in the following manner: $\Delta$ obeys the Klein-Gordon equation:
$$
(-\Box_x+m^2) \Delta=0
$$  
Taking advantage of the Poincar\'e symmetry, the angular integration can be trivialised and the Klein-Gordon operator transformed into the ordinary differential equation (see \eqref{box})
$$
-\Delta''-\frac{(D-1)}{\mu} \Delta'+m^2 \Delta=0
$$
Whose solutions are indeed Bessel functions. The correct linear combination can be found in different manners, for instance by imposing that $\Delta$ decays in the IR ($\left\vert \mu^2 \right\vert \to\infty$). 

\medskip
Consider now the slightly more complicated case of the one-loop corrected propagator $G_1$. We restrict our discussion to the massless field for simplicity. Calculations are usually carried--out in Fourier space, where
$$
G_1(k)=\Delta+\Delta \Sigma \Delta
$$
The one-loop self--energy $\Sigma$ in the modified minimal subtraction scheme  \cite{Srednicki:1019751} reads:
$$
\Sigma_{\overline{\textrm{MS}}}(k^2)=-\frac{\lambda^2}{12 (4\pi)^3} k^2+\frac{\lambda^2}{2(4\pi)^3} \int_0^1 dx D \ln(D/M^2)
$$
where $D=x(1-x)k^2$ and $M$ is the renormalization parameter. This expression is easily Fourier transformed to coordinate space and the IR behaviour is $G_1\propto \lambda^2 \mu^{-4} \ln \mu^2$. 
The same result can be retrieved immediately using again a differential equation formulation of the problem: $G_1(\mu)$ verifies 
$$
G_1(x,x')=\Delta(x,x')+\frac{\lambda^2}{2}\int dV_a dV_b \Delta(x,a)\Delta^2(a,b)\Delta(b,x') 
$$
Applying the massless Klein-Gordon operators on $x$ and $x'$, we obtain
$$
(-\Box_x)(-\Box_{x'})G_1(x,x')=(-\Box_x)\delta(x,x')+\frac{\lambda^2}{2}\Delta^2(x,x'). 
$$
The solution of which is\footnote{The local terms proportional to $\delta$ and its derivatives will only contribute by redefining the integration constants in different spacetime regions.}:
$$
G_1(\mu)=\frac{C_1}{\mu^2}+\frac{C_2}{\mu^4}+C_3\; \mu^2+C_4-c_6^2 \lambda^2 \; \frac{ 22+12\ln\mu^2}{576 \mu ^4},
$$
where $C_i$ are integration constants. The solution having the correct $\lambda \to 0$  behaviour and decaying in the IR reads:
$$
G_1(\mu)=\Delta(\mu)-c_6^2 \lambda^2 \; \frac{ 22+12\ln\mu^2}{576 \mu ^4}\sim_\textrm{IR}  -\frac{c_6^2 \lambda^2 \;\ln\mu^2}{48 \mu ^4} .
$$
We see already some advantages of the differential equation method in flat space, but the real advantages will become clear in the case of de Sitter spacetime. 

\medskip
We end this section by two remarks. First, we note that the dominant IR behaviour is included in the particular solution to the inhomogeneous equation, making it independent from any boundary conditions. This phenomena will turn out to be also true in the massless de Sitter case as well. Second, this differential equation method is applicable to large class of diagrams, but cannot be --at least immediately-- generalized  to all diagrams. 
 
\subsection{Interacting massive fields in de Sitter}

In this section we illustrate the differential equation method on the case of the \textit{massive} scalar field with a cubic self-interaction in de Sitter. The one-loop correction to the propagator for this theory has been calculated in the well-known paper \cite{Marolf:2010cd} using the de Sitter invariant Bunch-Davies vacuum and Watson-Sommerfeld transformations. The main result of that paper is that the 1-loop corrected propagator decays in the IR. We re-derive in this section this result using the differential equation technique.  

The free propagator, given in \ref{massivefreesolution}, can be be found by different methods. The most effective one being the differential equation method, as done for instance in \cite{Allen:1986wq}.

Now, applying the Klein-Gordon operator twice and taking full advantage of de Sitter invariance, we find that $G_1$ obeys:
$$
\mathbf{H}^2 G_1(z)=\frac{\lambda^2}{2}\Delta^2(z).
$$
This equation can be exactly solved, however we simply need here to extract the asymptotic IR behaviours, which is immediate. We find behaviours proportional to:
$$
\left\{z^\sigma,z^{\overline{\sigma}},z^{\sigma} \ln z, z^{\overline{\sigma}} \ln z, z^{2\sigma}\right\}
$$
in full-agreement with \cite{Marolf:2010cd}. In particular all possible IR behaviours are decaying in the IR. In order to implement the correct boundary conditions, one needs either the exact solutions (quite cumbersome), either a uniform approximation. We will study such approximations for the more interesting massless case in the rest of this paper. 

\medskip
The efficiency of our method is obvious in this last example. We also note the very interesting contribution from the same authors in the subsequent  paper \cite{Marolf:2011dj} where they prove that the massive field propagator is decaying in the IR at \textit{any loop} order, for any $n-$point function. The interested reader can also consult the closely related paper \cite{Hollands:2013gn}.

\section{Dyson--Schwinger equations}
\label{DSequations}
\begin{figure}[!htbp]
\begin{center}
\includegraphics[trim = 10mm 220mm 20mm 5mm, clip, scale=1]{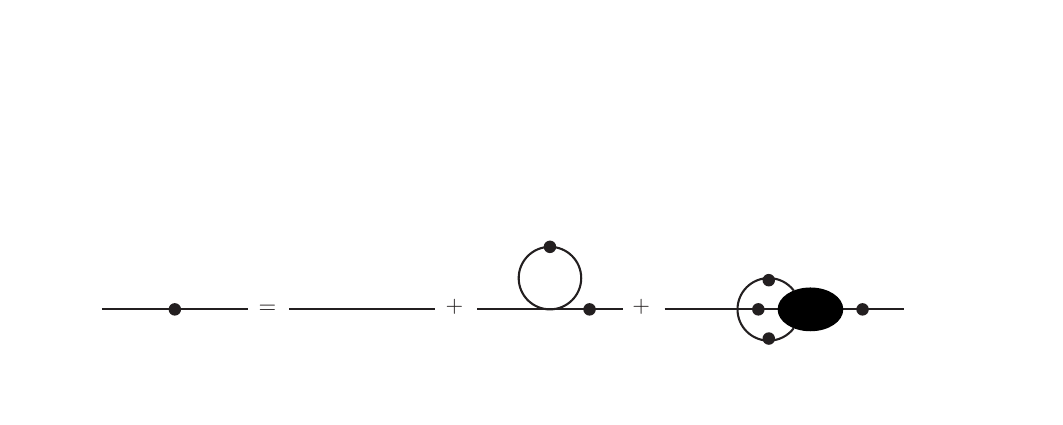}
\caption{\label{figSD}Dyson--Schwinger equation relating the two and four point functions. A blob on a propagator indicates that it is exact.}
  \end{center}
\end{figure}

The Dyson--Schwinger equations, an infinite set of integral equations between the n-point functions,  are the equations of motion of QFT. These equations, through some truncation schemes, give a convenient handle on non-perturbative effects in the theory. 
From now on we study the $4-$dimensional QFT given by the action:
$$
S=\int dV_x\left[\frac{1}{2} g_{\mu\nu}\partial^\mu \phi \partial^\nu \phi-\frac{1}{2} m^2 \phi^2-\frac{\lambda}{4!}\phi^4 \right],
$$
The (unrenormalized) Dyson--Schwinger equations for the exact propagator reads:
\begin{align}
 \label{Dysoneq}
(-\Box_x+m^2)G(x,x')=&-i \delta(x,x')-i \int dV_{y} \; \Sigma(x,y)G(y,x')\\
\nonumber\Sigma(x,y)=&\frac{1}{2}(-i \lambda)G(x,y)\delta(x,y)+\frac{1}{6}(-i \lambda)^2 \int dz\;G(x,z_1)G(x,z_2)G(x,z_3)\\
\nonumber &\Gamma_4(z_1,z_2,z_3,y),
\end{align}
where $dz=dV_{z_1}dV_{z_2}dV_{z_3}$, $G$ is the exact Feynman propagator, $\Delta$ the free one, $\Sigma$ the self-energy, which involves the exact four-point function $\Gamma_4$ and so on. A graphical representation of the first two DS equations is given in figure \ref{figSD}. The first equation is known as the Dyson equation which in flat spacetime is readily solved in Fourier space and is equivalent to summing the usual geometric series of self-energies: $G=\Delta/(1-\Delta \Sigma)$. The second equation, giving the self-energy, contains local and non-local contributions. We will study them both in what follows.

The Dyson--Schwinger equations  are an infinite tower of integral equations difficult to study without a heavy truncation scheme. One of the most frequently used schemes is the ladder--rainbow approximation, which consists of the replacement of the exact four-point function by its bare value: $\Gamma_4(z_1,z_2,z_3,y)\to -i\lambda\;\delta(z_1,y)\delta(z_2,y)\delta(z_3,y)$. After a quick discussion of the local contributions in section \ref{localsection}, we use the ladder--rainbow approximation throughout this work to study non-local contributions to the self-energy.

\medskip

\subsection{Local approximation and dynamical mass generation}
\label{localsection}
\begin{figure}[!htbp]
\begin{center}
\includegraphics[trim = 20mm 230mm 20mm 5mm, clip, scale=1]{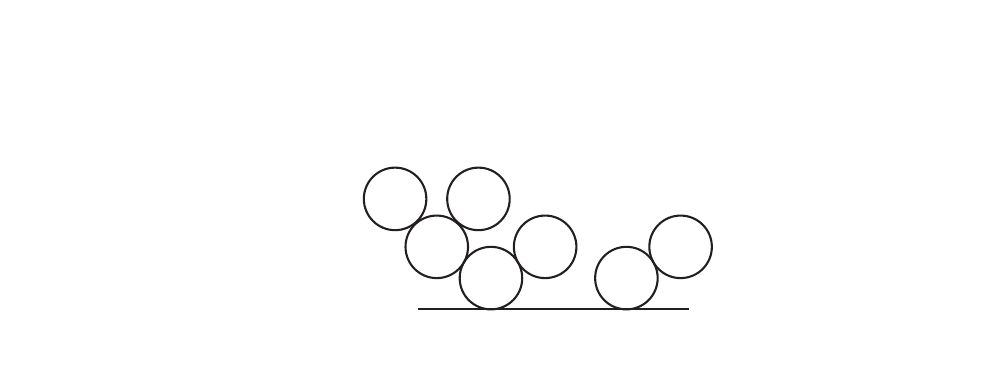}
 \caption{\label{figcactus}A typical cactus like diagram summed in the Hartree approximation.}
  \end{center}
\end{figure}

The local approximation of the self-energy, given by $\Sigma(x,y) = -i \dfrac{\lambda}{2} \;G(x,y) \delta(x,y)$ is the simplest approximation in which one can study the Dyson-Schwinger equations. It is known as the Hartree approximation an resumms the so--called cactus diagrams depicted in figure \ref{figcactus}. It leads to the equations:
\begin{align*}
(-\Box_x+m_{\textrm{dyn}}^2)G(x,x')=-i \delta(x,x'),\quad m_{\textrm{dyn}}^2=m_0^2+\frac{\lambda}{2} G(x,x)
\end{align*}
We take from now on the bare mass $m_0^2=0$. The first equation can be readily solved in terms of the hypergeometric function as in \eqref{massivefreesolution}. The second equation is the so--called gap equation and is explicitly given by:
$$-\lambda  \left(2 H^2-m^2_{\textrm{dyn}}\right) \left[\mathcal{H}_{\nu_{-}}+\mathcal{H}_{\nu_{+}}\right]+2 H^2\lambda -\left(\lambda +32 \pi^2\right) m^2_{\textrm{dyn}}=0
$$
where $\nu_{\pm}=\dfrac{1}{2}\left(1\pm\sqrt{9-\frac{4 m^2_{\textrm{dyn}}}{H^2}}\right)$ and $\mathcal{H}_z=\sum_{k=1}^\infty \left(\dfrac{1}{k}-\dfrac{1}{k+z} \right)$ is the harmonic number function. For small $\lambda$ the solution behaves like
$$
m^2_{\textrm{dyn}}\sim\frac{H^2\sqrt{3\lambda}}{4\pi}
$$
in agreement with \cite{Starobinsky:1994is,Serreau:2011gt} and numerous other works. 

\medskip
Several comments are in order. First, the resummed two--point function behaves now like a massive one and decay in the IR. This is our first result. Second, the dynamically generated mass $m^2_{\textrm{dyn}}$ vanishes in the flat spacetime limit ($H\to 0$) as expected. Finally, and perhaps the most important commentary concerning the dynamical mass generation: the actual result depends on how one defines the coincidence limit $G(x,x)$. This is already a non-trivial question in flat space, even if physically sound arguments strongly suggest that the tadpole diagram vanishes and no dynamical mass generation is possible. The situation in de Sitter is at least as intricate (see appendix C of \cite{Marolf:2010cd} and the recent detailed analysis in \cite{Nacir:2013xca}).

We close this section by noting that in the literature, the gap equation is only obtained for small masses and coupling. On the contrary, we have obtained its general form. As a consequence, we are also able to study its solution in the strong coupling regime, $\lambda\to \infty$. Interestingly, we find in this limit that $m^2_{\textrm{dyn}}\to 2H^2$, which corresponds to the conformally invariant field, whose two--point function simply reduces to $\left[16\pi^2 (1-z)\right]^{-1}$. To our knowledge, this strong coupling limit has not been given before and the precise meaning of the appearance of the conformal field is yet unclear to us.

\subsection{Non-local approximation: Ladder-Rainbow}
\label{nonlocalsection}

From now on we will discard eventual local contributions to the self-energy and concentrate on the non-local ones. First, let us  analyze the Dyson--Schwinger equations for our purposes in some detail.

Since we discard local corrections, the expansion of the self-energy starts at two loops, with the first term in the expansion indeed furnishing a primitive element in
the Hopf algebra $H$ of $\phi^4$ theory, which provides a Hochschild one-cocycle $B_+$. Omitting higher Hochschild co-cycles -which seems reasonable as they do not alter the algebraic structure
of the Dyson--Schwinger equations-, we find the combinatorial Dyson--Schwinger equation for the self-energy 
$$
X(\lambda^2)=1-\frac{\lambda^2}{6}B_+\left(\frac{1}{X^3(\lambda^2)}\right).
$$ 
Its fix-point is a formal series $\in H[[\lambda^2]]$ such that the application of renormalized Feynman rules
$\Phi_R$ gives $\Phi_R(X(\lambda^2))=1-\Sigma$.

Hochschild cohomology ensures that we renormalized by local counterterms, and ensures that each contributing graph is divided by its symmetry factor, as it should \cite{Kreimer:2010tn,Broadhurst:2001upa}.

\begin{figure}[!htbp]
\begin{center}
\includegraphics[trim = 20mm 220mm 20mm 5mm, clip, scale=1]{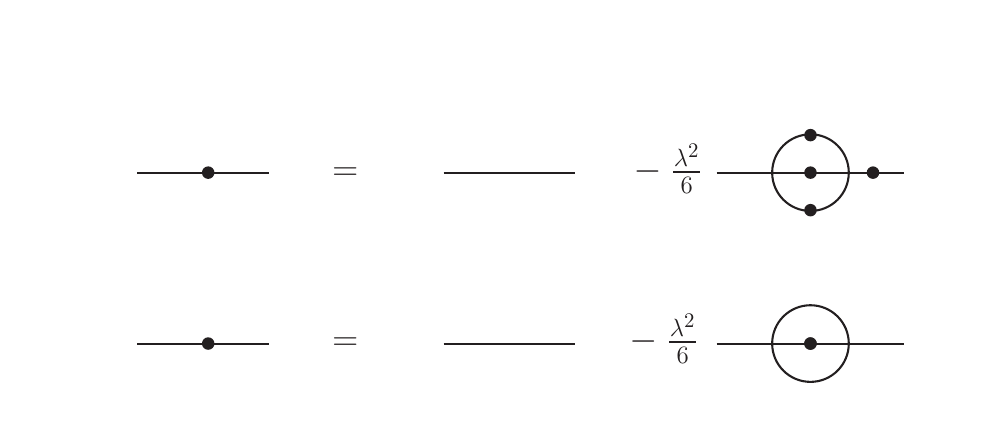}
        \caption{\label{fignlSD}Non-linear and linear Dyson--Schwinger equations in the ladder-rainbow approximation.}
  \end{center}
\end{figure}

We drastically simplify this system by linearizing it to a commutative and co-commutative Hopf algebra generated from simple concatenations of the cocycle $B_+$.
$$
X(\lambda^2)=1-\frac{\lambda^2}{6}B_+\left(X(\lambda^2)\right).
$$ 
Graphically, the two equations read as in Figure \ref{fignlSD}.

\begin{figure}[!htbp]
\begin{center}
\includegraphics[trim = 20mm 235mm 10mm 5mm, clip, scale=1]{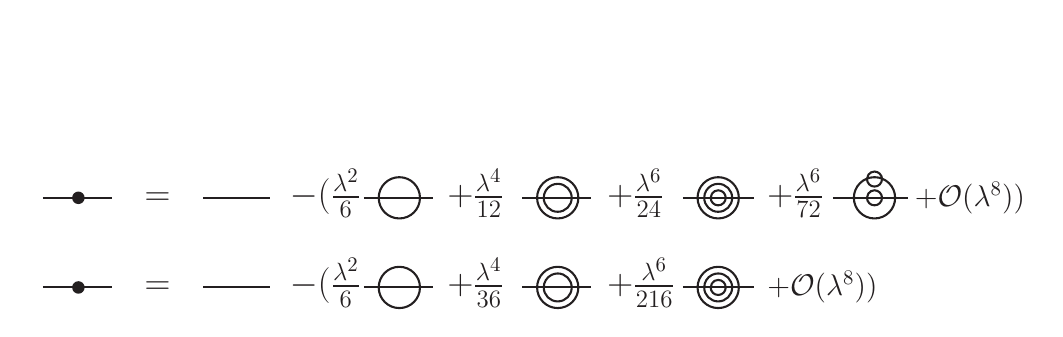}
\caption{\label{figpSD}Non-linear and linear Dyson--Schwinger equations (both in the ladder--rainbow approximation) expanded out in perturbation theory. Note that the linear Dyson--Schwinger equations misses some graphs, which also results in wrong symmetry factors. It nevertheless gives a strict subset of the full expansion at any order.}
  \end{center}
\end{figure}

\newpage
Explicitly, the two expansions in graphs read as in Figure \ref{figpSD}.

Note that in the linearized case, graphs do not contribute by their symmetry factors any more, as ``insertion places" are missing. Explicitly, the Dyson--Schwinger equation in the linearized ladder--rainbow approximation reads (see \cite{Delbourgo:1996nw} for the flat space case):
$$
(-\Box_x+m^2)(-\Box_{x'}+m^2)G(x,x')-\frac{\lambda^2}{6}  \Delta^2(x,x') G(x,x')=-i (-\Box_x+m^2)\delta(x,x').
$$ 
For maximally symmetric spacetimes and vacuum states this reduces to:
\begin{equation}
\left[(-\Box+m^2)^2-\frac{\lambda^2}{6}  \Delta^2(x,x') \right] G(x,x')=-i (-\Box+m^2)\delta(x,x').
\end{equation}
In terms of the $z$ variable and the hypergeometric operator $\mathbf{H}$, using the formulas \eqref{box}, the homogeneous \footnote{The inhomogeneity proportional to a delta function will be responsible to a shift that can be absorbed in the integration constants.} part of the preceding equation becomes:
\begin{equation}
\label{mastereq}
\left(\mathbf{H}^2-\epsilon\: \Delta^2\right)G(z)=0, \quad \epsilon=\lambda^2/6.
\end{equation}
Note that for the non--linear Dyson--Schwinger equation, the corresponding differential equation would become highly non-linear even in flat space --see section \ref{nonlinearladderrainbow}--,
and on top of that non-linearity we would have all the complexities of the hypergeometric operator on which we now focus. 

This ordinary differential equation is the central equation of our work and we will refer to it as the master equation in what follows.
We will use the following boundary conditions, which are the natural generalization of the  Bunch--Davies vacuum state: regularity for antipodal points (i.e at $z=0$) and a flat spacetime singularity at short distances (at $z\to 1$). The flat space singularity is proportional to  (see Appendix \ref{appflat}):
\begin{equation}
\label{flatUV}
\mu^{-\frac{1}{2}
   \sqrt{\frac{\sqrt{\epsilon +64
   \pi ^2}}{\pi ^2}+8}}.
\end{equation}
Developing the relation $z=\cos^2 \left( \frac{\mu}{2R}\right)$ near $\mu \to 0$, we have the asymptotic behaviour of $G$ at $z=1$:
$$
G \under_{z\to 1}  \: \frac{1}{16\pi^2}  \: (1-z)^{\nu}, \quad \nu=-\frac{1}{4} \sqrt{\frac{\sqrt{\epsilon +64 \pi^4}}{\pi ^2}+ 8}.
$$

\section{Asymptotics of the master equation}
\label{asymptoticofmaster}

\subsection{Perturbation series and leading logarithms}
First, let us tackle the master equation perturbatively in $\epsilon$. This is relevant to understanding the physics of the problem as well as the need of a more refined asymptotic analysis. Writing $
G=G_0+\epsilon \; G_1 + \epsilon^2 \;G_2+\cdots$,
it is relatively easy --although tedious-- to compute the first terms in this series. The result involves logarithmic and polylogarithmic functions.

A very interesting fact worth mentioning at this level is that, because we work at the level of the differential equation, no IR divergencies appear whatsoever: the differentiation serves as a canonical and non--ambiguous regularization procedure. These facts will be discussed in much more detail in the forthcoming publication. Instead of divergencies, we obtain perfectly well--defined expressions for every $G_n$, and they grow in the IR. More precisely, we have the ``leading logarithms":
$$
G_0(z)\sim -\frac{\ln z}{8\pi^2}, \quad G_1(z)\sim-\frac{\ln^5 z}{92160\pi^6}, \cdots
$$  
Actually we are able to obtain by induction the explicit leading large $z$ behaviour for an arbitrary order:
$$
G_n(z) \sim -\frac{\Gamma(5/4)}{2^{3+10n} 9^n \pi^{2+4n}\Gamma(1+n)\Gamma\left(n+5/4\right)} \ln^{1+4n}z.
$$
It is of course very tempting to sum up these leading logarithms, which we do. The result is convergent but wrong, i.e the sum of the sub--leading terms in $G_n$ are not negligible. We note that the exact same phenomenon occurs for the quantum anharmonic oscillator \cite{Bender:1996vf}. We thus need a more powerful method to derive the asymptotics of the master equation. We will use here the WKB method.

\subsection{WKB approximation}
We study the homogeneous master equation \eqref{mastereq}, which we reproduce here explicitly for convenience:
\begin{align}
\label{expeq}
&G^{(4)}+\left(\frac{6}{z}+\frac{6}{z-1}\right)G^{(3)}+\left(\frac{6}{z^2}-\frac{22}{z}+\frac{22}{z-1}+\frac{6}{(z-1)^2}\right)G''\\
\nonumber&+\left(\frac{8}{(z-1)^2}-\frac{8}{z^2}\right) G'- \frac{\epsilon}{256 \pi ^4(1-z)^2 z^2} \left(\frac{1}{1-z}-2 \ln(1-z)\right)^2 G=0.
\end{align}
This equation  possesses  one regular singular point at $z=0$ and two irregular singular points at $z=1$ and $z=\infty$. The local study performed around these points in Appendix \ref{appendixloc} shows that the interaction term (proportional to $\epsilon$) is negligible near $z=0$ and is very important near $z=\infty$.

We then use a WKB approximation (or Liouville-Green for mathematicians) to obtain global approximations to the four fundamental solutions of \eqref{expeq}. The solution that posses the correct $z\to1$ has the following IR behaviour:

$$
\boxed{G\sim\frac{1}{z^{3/2} \ln^{\frac{3}{4}}(-z)} \exp \left[ -\left( \dfrac{\sqrt[4]{\epsilon } \ln^{\frac{3}{2}}(-z)}{3 \sqrt{2}\pi }+\dfrac{9 \pi  \sqrt{\ln(-z)}}{\sqrt{2} \sqrt[4]{\epsilon}}\right)\right].}
$$

As the other solutions have a sub-leading contribution near $z=1$, they (notably the one with $\omega=1$ in the appendix \ref{appendixloc}) can spoil the IR behaviour by making it grow instead of decaying. The absence of this solution will be checked via a rigorous analysis elsewhere.  

We note that, while the approximation near $z=1$ obtained via a certain WKB approximation is only valid for strong couplings $\epsilon \ll 1$, the asymptotic behaviour obtained via local analysis (see Appendix \ref{appendixloc} ) is exact for any $\epsilon$.

Finally an important remark: the phenomenon described here belongs to a large class of phenomena in mathematics and physics known as secular perturbation theory. The simplest such example is given by the boundary value problem \cite{Holmes:2012uo}:
\begin{align*}
y''+\epsilon y'+y=0, \quad \textrm{for } t>0, \quad y(0)=0, y'(0)=1. 
\end{align*}
To the first order in perturbation theory in $\epsilon$ the solution is unbounded for large $t$:
$$
y(t)=\sin(t)-\frac{1}{2} \epsilon \; t \sin(t).
$$
Unlike this perturbative solution, the exact solution is bounded:
$$
y(t)=\frac{1}{\sqrt{1-\epsilon^2/4}} e^{-\epsilon t/2} \sin\left(t\:\sqrt{1-\epsilon^2/4}\right).
$$ 
In the case under consideration, the multiple scales method \cite{Bender:294110,Holmes:2012uo} (consisting of introducing a new scale $\tau=\epsilon t$, treat it as independent of $t$, and use this freedom to kill the secular terms) yields, to the first iteration, the uniform approximation
$$
y(t)\sim e^{-\epsilon t/2} \sin(t).
$$
We will investigate the applicability of these multiple scales methods to the study of strong IR effects in de Sitter in a future publication. 

%Such secular behaviour is usually related to a resonance in the system. Accordingly, we suggest that the expansion of the de Sitter background plays the role of a resonant external coupling. 

\section{Non--linear ladder--rainbow approximation}
\label{nonlinearladderrainbow}
In the non--linear ladder--rainbow approximation the two--point function verifies:
$$
\mathbf{H}^2 G-\epsilon\; G^3=0. 
$$ 
This equation resumms all of the diagrams shown in the first equation of fig. \eqref{figpSD}. Note that it does not take into account repeated chains of self-energy. The large $\mu,z$ asymptotics of the linear equation is of course more involved than the linear case. However we are able to make some progress in the flat space case by tackling the problem perturbatively in $\epsilon$. Note that the full equation flat space equation, taking into account chains have been studied in \cite{Broadhurst:2001upa} for the cubic interaction in flat space.

We are able to find the explicit form of the leading infrared term for an arbitrary perturbation order. More precisely we have in the IR:
\begin{align}
\nonumber G(\mu)&=G_0(\mu)+\epsilon \; G_1(\mu)+\epsilon^2 \; G_2(\mu)+\cdots \\
G_n(\mu)&\sim \frac{(-1)^{3n}\; \nonumber \Gamma\left(\frac{1}{2}+n\right)}{8^{1+3n}\; \pi^{\frac{5}{2}+4n}\;\Gamma(1+n)} \frac{\ln^{n} \mu}{ \mu^2}. 
\end{align}
These leading behaviours can be resummed and lead to 
$$
G\sim  \frac{2 \sqrt{2}}{\mu ^2 \sqrt{\epsilon  \ln \mu}}.
$$
Whether one can neglect the sum of the sub--leading terms is a crucial question whose investigation is beyond the scope of this paper.
\section{Discussion}
\label{discussion}

We end our work with a concise discussion of several important issues.

\medskip 
A crucial technical point in our analysis is that we are able to take full-advantage of de Sitter-symmetry by transforming the Dyson--Schwinger integral equation into an ordinary differential equation (depending only on the scalar variable $z$). This not only drastically simplifies the computations, but it also constitutes a non--ambiguous  IR regularization procedure.

\medskip
We have successfully resummed non--local contributions to the self-energy. To our knowledge, this is the first time such results are obtained. It is however important to stress that our conclusions were drawn in a very restrictive context, namely the ladder--rainbow approximation, which is a drastic simplification of the full theory.  A better control of this approximation has to be achieved, mainly by summing over a larger family of diagrams, for instance using the non-linear Dyson--Schwinger equation discussed above. Another interesting path to follow is to combine insights from our present work and from the ``physical momentum representation" of de Sitter correlators  developed in \cite{Serreau:1516680,Parentani:2013dr}. The aim would be to resum the Dyson series of the ladder-rainbows self-energies. 

We also note that the difficulty to rigorously justify the used approximation is a general trend in hard problems such as non-perturbative QFT, a classical example being the study of bound-states using the Bethe-Salpeter equation \cite{Itzykson:100772}. 

\medskip 
We have been able to avoid any discussion about UV-renormalization because we transform the integration over loops into  differential equations. Hence renormalization only intervenes through the integration constants. This is actually the generic situation for the study of solutions of Dyson--Schwinger equations: for a kinetic renormalization scheme, renormalization condition amount to fixing the boundary conditions of the equations.

\medskip The secularity of perturbation theory exists already in flat space. However, unlike the de Sitter case, it does not change the IR behaviour of the two--point function, because of a rapidly decaying overall $\mu^{-2}$ factor. Indeed the ladder--rainbow two--point function in flat space reads (see appendix \ref{appflat}):
$$
\mu^{-\sqrt{2 +\frac{\sqrt{\epsilon +64 \pi^4}}{4 \pi ^2}}}=\frac{1}{\mu^2} \left[1-\frac{\epsilon  \ln \mu}{256 \pi ^4}+\frac{\epsilon ^2 \left(5 \ln\mu+2 \ln^2\mu\right)}{262144 \pi^8}+\cdots\right]. 
$$

\medskip
In non--stationary situations, one should use the so-called Schwinger--Keldysh (also known as the in-in) QFT formalism. However 
de Sitter spacetime is particular in that the in-in and the Euclidean formalisms are equivalent. This has been proven for the massive interacting case in \cite{Higuchi:2011cb}. Whether this equivalence holds in the massless limit likely depends on how one interprets and treat the strong IR effects that arise in this case, as discussed in the introduction. This is certainly an interesting point to study.

\medskip
The IR behaviour of the graviton field in de Sitter spacetime is one of the most important open questions in cosmology. We believe that the non-perturbative effects exhibited in the present work for the mmc scalar field can illuminate this issue. On top of these non-perturbative effects, we also expect an interesting interplay with non-trivial gauge artifacts similar to the ones already exhibited for the photon field in de Sitter in  \cite{Youssef:2011eh}. 

\medskip
As already mentioned in the introduction, we neglect the back--reaction of the quantum fields on the background metric. However the fact that the interacting two--point function $G$ decays in the IR is of paramount relevance to the difficult but crucial issue of whether the back--reaction can be neglected to begin with. Ultimately, this is equivalent to understanding the stability of the de Sitter spacetime.

\begin{acknowledgments}
D. Kreimer is supported by the Alexander von Humboldt Foundation and the BMBF through an Alexander von Humboldt Professorship.
\end{acknowledgments}

\newpage

\appendix

\section{Conventions}
\label{conv}
Here we list our kinematical conventions. The metric is the mostly plus one. Consider the action
$$
S=\int d^4x \sqrt{-g}\left[\frac{1}{2} g_{\mu\nu}\partial^\mu \phi \partial^\nu \phi-\frac{1}{2} m^2 \phi^2-\frac{\lambda}{4!}\phi^4 \right] 
$$
where $g$ is the determinant of the metric. The free Feynman propagator $\left\langle T \phi(x)\phi(x')\right\rangle=\Delta(x,x')$ verifies:
$$
(-\Box_x+m^2)\Delta(x,x')=-i \frac{\delta(x-x')}{\sqrt{-g}}.
$$
Feynman rules are given by: 
\begin{align*}
\textrm{Propagator:} &\quad \Delta(x,x') \\
\textrm{Vertex:} &\quad -i \lambda \\
\end{align*}

\section{Flat space master equation}
\label{appflat}
In the flat space the master equation \eqref{mastereq}, with $A=1/\mu$  and $\Delta(\mu)=\frac{1}{4\pi^2 \mu^2}$
is readily solved and gives the four fundamental solutions
with 
$$
\mu^{\rho}, \quad \rho =\pm \sqrt{2 \pm \frac{\sqrt{\epsilon +64 \pi^4}}{4 \pi ^2}}.
$$
The boundary conditions, included in the integral equation for instance, select the solution for which $\rho=-\sqrt{2 +\frac{\sqrt{\epsilon +64 \pi^4}}{4 \pi ^2}}$.

\section{Local analysis}
\label{appendixloc}
We study the homogeneous master equation \eqref{mastereq}, which we reproduce here explicitly for convenience:
\begin{align}
\label{expeq}
&G^{(4)}+\left(\frac{6}{z}+\frac{6}{z-1}\right)G^{(3)}+\left(\frac{6}{z^2}-\frac{22}{z}+\frac{22}{z-1}+\frac{6}{(z-1)^2}\right)G''\\
\nonumber&+\left(\frac{8}{(z-1)^2}-\frac{8}{z^2}\right) G'- \frac{\epsilon}{256 \pi ^4(1-z)^2 z^2} \left(\frac{1}{1-z}-2 \ln(1-z)\right)^2 G=0.
\end{align}
This equation  possesses  one regular singular point at $z=0$ and two irregular singular points at $z=1$ and $z=\infty$. We perform here a local analysis near these points.

\subsection{Frobenius series at $z=0$}
The indical polynomial is $P_0(\lambda)=\lambda^2(\lambda^2-1)$. Our equation being fourth-order and the roots of  $P_0$ being separated by integers, some care has to be taken in the application of the Frobenius method (see \cite{Barkatou:2010ta} for the explicit algorithm). We obtain the local behaviours:
$$
\left\{\frac{1}{z},\ln z, 1, z  \right\}
$$
More precisely the first terms of the four solutions are given by:
\begin{equation*}
\left\{
\begin{array}{rl}
G_1(z) &\sim z+\dfrac{2z^2}{3}+\left(\dfrac{\epsilon }{18432\pi^4}+\dfrac{5}{9}\right)z^3\sim z,\\
&\\
G_2(z) &\sim \dfrac{1}{z}+84-\left(216+\dfrac{15\epsilon}{512\pi^4} \right)z+\left[ -48+\left(36+\dfrac{3 \epsilon}{256\pi^4} \right)z\right]\ln z \sim \dfrac{1}{z},\\
&\\
G_3(z) &\sim 1-z+\left(-\dfrac{2}{3}+\dfrac{\epsilon}{3072\pi^4} \right)z^2\sim 1,\\
&\\
G_4(z) &\sim 9z+\left( \dfrac{40}{9}-\dfrac{7\epsilon}{4608\pi^4}\right)z^2+\left[2-2z+\left(-\dfrac{4}{3}+\dfrac{\epsilon}{1536\pi^4} \right)z^2 \right] \ln z \sim 2 \ln z.
\end{array} \right.
\end{equation*}  
This analysis implies that, the boundary condition imposing regularity  at $z=0$ eliminates $2$ of the $4$ solutions.

\subsection{Asymptotic behaviour near $z=1$}
\paragraph{$\epsilon=0$.}

In this case the point $z=1$ is a regular singular point. The free master equation being invariant under the transformation $z\to 1-z$, the asymptotic behaviours at $z=1$ can be directly obtained from the behaviour near $z=0$ (also the equation is then exactly solvable) and are given by:
$$
\left\{1-z,\frac{1}{1-z},\ln(1-z),1\right\}.
$$

\paragraph{$\epsilon>0$.}
Even if the point $z=1$ is an irregular singular point, this ``irregularity" appears only at the first order when $z\to 1$. At zeroth-order in $1-z$, one can thus still define an indicial polynomial which is given by:
$$
P_1(\lambda)=\lambda ^4-\lambda ^2-\frac{\epsilon }{256 \pi^4}. 
$$
The roots of which read:
$$
\lambda=\pm \frac{1}{4} \sqrt{\frac{\sqrt{\epsilon +64 \pi^4}}{\pi ^2}\pm 8}
$$
and are not separated by integers for generic $\epsilon$. The leading behaviours near $z=1$ is thus given by 
$$
(1-z)^{\lambda}, \quad \lambda=\pm \frac{1}{4} \sqrt{\frac{\sqrt{\epsilon +64 \pi^4}}{\pi ^2}\pm 8}
$$
We can verify that this is the full leading behaviour at $z=1$.
\subsection{Asymptotic behaviour near $ z \to  \infty$}
By making the ansatz $G=e^S$ and using the dominant balance method recursively \cite{Bender:294110}, we are able to derive the leading behaviour near infinity:
$$
G_\omega\sim\frac{1}{z^{3/2} \ln^{\frac{3}{4}}(-z)} \exp \left[ \left( \dfrac{\omega \sqrt[4]{\epsilon } \ln^{\frac{3}{2}}(-z)}{3 \sqrt{2}\pi }+\dfrac{9 \pi  \sqrt{\ln(-z)}}{\sqrt{2} \omega \sqrt[4]{\epsilon}}\right)\right], \quad \omega^4=1.
$$
Obtaining this result, necessitates some lengthy calculations that we choose not to reproduce here. 
\newpage

\newpage

\section{WKB approximation}
Consider the general fourth-order differential equation (of course the parameter $\delta$ can be introduced in many different ways, leading to more or less accurate approximations):
$$
\delta^4 \left[ G^{(4)}+ a_3 G^{(3)}+ a_2 G''+  a_1 G' \right]- a_0 G=0. 
$$
When $\delta\to 0$, each of the four elementary solutions in the WKB approximation read (in our situation, $\delta=\epsilon^{-1/4}$, meaning that we perform a strong coupling expansion):
$$
G\sim \exp\left[ \frac{1}{\delta} \left(S_0+\delta S_1+\delta^2 S_2 \right)       \right]
$$
where
\begin{align}
\label{WKBgeneral}
\nonumber S_0&= \omega  \int^z \sqrt[4]{a_0}, \qquad \omega^4=1, \\
S_1&=\frac{-1}{8} \left(3 \log a_0+2 \int^z
   a_3 \right),\\
\nonumber S_2&= \frac{1}{128 \omega} \int^z \frac{1}{\sqrt[4]{a_0}} \left[
-\frac{45 a_0'{}^2}{a_0{}^2}+48 a_3'+\frac{40
   a_0''}{a_0}+12 a_3{}^2-32 a_2
\right].
\end{align}
These solutions have the asymptotic behaviours
\begin{equation*}
\left\{
\begin{array}{rl}
G_\omega &\under_{z \to 1} \; (1-z)^{\dfrac{\omega  \sqrt[4]{\epsilon }}{4 \pi }+\dfrac{\pi }{\omega  \sqrt[4]{\epsilon }}}, \\
&\\
G_\omega  &\under_{z\to \infty}  \dfrac{1}{z^{3/2} \ln^{\frac{3}{4}}(-z)} \exp \left[\left( \dfrac{\omega \sqrt[4]{\epsilon } \ln^{\frac{3}{2}}(-z)}{3 \sqrt{2}\pi }+\dfrac{9 \pi  \sqrt{\ln(-z)}}{\sqrt{2} \omega \sqrt[4]{\epsilon}}\right)\right]. 
\end{array} \right.
\end{equation*}  
Comparing the WKB results to the exact local analysis performed before, we find that they are exactly the same as $z\to \infty$. They also coincide near $z \to 1$ in the strong coupling limit $\epsilon \gg 1$. This is a verification of our previous calculations. The ``minimal" choice corresponding to the boundary value at $z=1$ will be $\omega=-1$, implying that the real part of the resummed two--point function decay at $z\to  \infty$.

However this WKB approximation breaks down near $z=0$, preventing us from implementing the boundary conditions at $z=0$. This breakdown can be explicitly seen because the following relations  in the expansion $G\sim \exp\left[ \frac{1}{\delta} \left(S_0+\delta S_1+\delta^2 S_2 \right)       \right]$ do not hold near $z=0$:
$$
S_0 \gg S_1 \gg S_2.
$$
This was also expected, since the point $z=0$ is a degenerate regular singular point, i.e having equal exponents (see \cite{Olver:1974th} p.$202$). The global analysis of such equations, specially higher-order ones, is a non-trivial task (see \cite{Koike:2000vz} and references therein) that we will examine in a future work.

\bibliography{Bibliography}

\end{document}